# *In Situ* Measurements of Stress-Potential Coupling in Lithiated Silicon

V.A. Sethuraman,[1,z] V. Srinivasan,[2,z] A.F. Bower,[1,z] P.R. Guduru[1,*]

[1] Division of Engineering, Brown University, Providence, Rhode Island 02912, USA
[2] Lawrence Berkeley National Laboratory, Berkeley, California 94720, USA

[z] vj@cal.berkeley.edu; VSrinivasan@lbl.gov; Allan_Bower@Brown.edu; Pradeep_Guduru@Brown.edu

An analysis of the dependence of electric potential on the state of stress of a lithiated-silicon electrode is presented. Based on the Larché and Cahn chemical potential for a solid solution, a thermodynamic argument is made for the existence of the stress-potential coupling in lithiated-silicon; based on the known properties of the material, the magnitude of the coupling is estimated to be *ca.* 60 mV/GPa in thin-film geometry. An experimental investigation is carried out on silicon thin-film electrodes in which the stress is measured *in situ* during electrochemical lithiation and delithiation. By progressively varying the stress through incremental delithiation, the relation between stress change and electric-potential change is measured to be 100 – 120 mV/GPa, which is of the same order of magnitude as the prediction of the analysis. The importance of the coupling is discussed in interpreting the hysteresis observed in potential *vs.* state-of-charge plots, and the role of stress in modifying the maximum charge capacity of a silicon electrode under stress.

## 1. INTRODUCTION

Silicon is considered to be a promising anode material to increase the energy density of state-of-the-art lithium-ion batteries. The fully lithiated phase of silicon at room temperature is $Li_{3.75}Si$, which translates to a maximum theoretical capacity of 3579 mAh g$^{-1}$,[1,2] much higher than that of graphite (372 mAh g$^{-1}$).[3,4] When used in a battery, the high capacity can result in a significant increase in the energy density and specific energy of the cell (by as much as 35%),[5] and could help lower the cost per kWh. The high capacity combined with the low discharge potential and low cost makes silicon an attractive choice for use as negative electrodes in lithium-ion batteries. A large body of literature exists on the electrochemical and mechanical performance of anodes made of pure silicon and composites in which silicon is one of the constituents.[6] Electrochemical lithiation and delithiation of silicon at ambient temperature has been extensively studied in recent years in such forms as nanowires,[7-9] amorphous thin films,[11-16] crystalline thin films,[17] crystalline powder,[18,19] composites,[20-23] mixtures with other metals,[24-28] and mixtures with carbon.[29-35] Silicon-graphite composite anodes are getting closer to commercialization,[36] with companies such as Mitsui Mining announcing the commercialization of composite anodes based on silicon particles coated with copper, that show excellent cycle life and capacity retention.[37]

Regardless of the nature of the silicon anode geometry studied, the electrochemical-lithiation potential *vs.* Li/Li$^+$ at a given state-of-charge (SOC) is considerably lower than the electrochemical-delithiation potential *vs.* Li/Li$^+$ at that SOC. Furthermore, in silicon, the lithiation and delithiation phenomena at practical rates occurs in the Tafel regime[38,39,40] and as a consequence, the difference between lithiation and delithiation potentials at a given SOC, defined as the potential offset, appears to be nearly rate independent.[7,41] For example, even at low rates (*e.g.* C/10), the potential *vs.* Li/Li$^+$ during lithiation is lower than during delithiation by approximately 320 mV.[18] Data in the literature shows that the offset exists irrespective of particle size,[42] and in nanowires,[7] thin films[13,43] and composite electrodes,[18] with values ranging from 320 mV to 250 mV. As a consequence of the potential offset, the silicon electrode exhibits a hysteresis loop (in a potential *vs.* capacity plot) upon lithiation followed by delithiation or *vice versa*. The hysteresis loop created during a lithiation and delithiation cycle is not sufficient to define the state of the Li$_x$Si system, even at low rates. Also, the potential obtained at any SOC depends on the cycling history of the Li$_x$Si system and therefore, cannot be used as an indication of the SOC of the cell.[44,45] The area of the hysteresis loop gives the energy dissipated in a single charge-discharge cycle. Hence, the potential offset is a measure of the loss of the cell efficiency (*e.g.*, 91.5% energy efficiency at low rates when a silicon anode is paired with a commercial 3.8 V LiNi$_{0.8}$Co$_{0.15}$Al$_{0.05}$O$_2$ cathode). Traditionally, the energy dissipation is thought to occur at the electrode/electrolyte interface (because of the over-potential required to drive a current) and IR losses elsewhere in the battery. However, the sources of the potential offset in lithiated silicon are neither fully understood nor satisfactorily explained thus far. Reports in the literature suggest that the hysteresis could be thermodynamic,[46] caused by a sluggish kinetics,[38,39] and possibly due to breaking of Si bonds.[47] One of the objectives of this study is to understand the role of electrode mechanics in contributing to the potential hysteresis.

It is well recognized that the large volume expansion of alloy anodes such as silicon leads to very high strains, possibly leading to fragmentation and capacity fade.[48,49] Sethuraman *et al.*[50] measured *in situ* stress evolution in silicon thin-film electrodes during electrochemical cycling, and reported that the film undergoes alternative compressive and tensile plastic deformation during lithiation and delithiation, respectively. Also, the stress magnitude was seen to be as high as 1.5 GPa. Although the effects of such high stress on the mechanical integrity of the anodes have been investigated by many researchers, its influence on the electric potential of the anode has received little attention. The next section presents a thermodynamic argument based on Larché and Cahn chemical potential[51-54] for the existence of such coupling and estimates its magnitude from the known properties of lithiated silicon. It is followed by a section which presents an experimental investigation to measure the magnitude of the stress-potential coupling, in which the film is delithiated in small increments to cause a large change in stress, while measuring the electrode potential. The implications of the stress-potential coupling in silicon electrodes is then discussed in the context of interpreting the hysteresis loops observed in the potential-charge plots during electrochemical cycling as well as the maximum realizable charge capacity of lithiated-silicon electrodes.

## 2. CHEMICAL POTENTIAL OF A SOLID-SOLUTION UNDER STRESS

The silicon thin films considered in this work are fabricated by sputtering process, which results in amorphous films. It is assumed that electrochemical lithiation of an amorphous silicon film results in an amorphous interstitial solid solution of silicon and lithium.[1,2,15] As the Si electrode is lithiated and delithiated, we assume that the film remains as an amorphous solid solution whose solute (Li) concentration varies during electrochemical cycling. We choose to describe the chemical potential of lithium in the solid solution with the Larché-Cahn potential[51-54] as shown below:

$$\mu_{Li} = \mu_{Li}^0 + RT \log\left(\gamma \frac{c}{c_{max} - c}\right) - \frac{v_{Si}\eta}{3}\sigma_{kk} - \frac{v_{Si}\beta_{ijkl}}{2}\sigma_{ij}\sigma_{kl} \qquad 1$$

where $\mu_{Li}$ is the chemical potential of Li atoms in the Li-Si alloy (per mole of Li atoms); $c$ is the Li concentration, defined as the ratio of the number of atoms of Li to that of Si in a unit volume; $c_{max}$ is the maximum possible value of $c$; $\mu_{Li}^0$ is a reference chemical potential corresponding to a reference concentration of $c = c_{max}/2$ and a stress-free state; $\gamma$ is the activity coefficient which takes a value of unity when $c = c_{max}/2$; $\eta$ is the rate of change of volumetric strain ($\varepsilon_v^c$) due to lithiation, defined as $\eta = d\varepsilon_v^c / dc$; $v_{Si}$ is the molar volume of silicon (volume per mole of Si atoms before alloying with Li), $\sigma_{ij}$ is the stress tensor, and $\beta_{ijkl} = ds_{ijkl}/dc$, where $s_{ijkl}$ is the elastic compliance tensor. The assumptions and limitations of the above description for the chemical potential should be clearly understood. It is meant to describe the effects of stress and alloy composition in the context of small strains, i.e., the strains induced by the combined effects of alloying, and mechanical loading are small enough that additive decomposition of the small-strain tensor $\varepsilon_{ij} = \varepsilon_{ij}^c + \varepsilon_{ij}^e + \varepsilon_{ij}^p$ is valid, where $\varepsilon_{ij}^c$ is the strain due to lithium concentration alone (also referred to as compositional strain), $\varepsilon_{ij}^e$ is the elastic strain, and $\varepsilon_{ij}^p$ is the plastic strain.

However, in case of the Si-Li alloy, the volumetric strain can be as large as 270% when $c \sim 3.75$.[55] To be accurate, one would have to develop a new form of the chemical potential using finite kinematics and an appropriate conjugate pair of stress and strain measures in writing the Gibbs free energy (e.g., Lagrangian strain and material stress). The errors arising from neglecting the finite kinematics can be substantial, especially at high Li concentrations. Although we recognize these limitations, as a starting point, we choose to proceed with equation 1 as an approximation. Moreover, for small Li concentrations, when the volume expansion is still within the small-strain approximation, equation 1 would be a good description of the chemical potential. Further, Larché and Cahn assume small solute concentrations, which is clearly violated in the Si-Li system in which the ratio of the number of solute atoms to that of solvent atoms can be as high as 3.75. An additional assumption in writing equation 1 is that the compositional strain is isotropic, i.e., $\varepsilon_{ij}^c = (\varepsilon_v^c/3)\delta_{ij}$, where $\delta_{ij}$ is the Kronecker delta.

## 2.1 ESTIMATION OF STRESS-POTENTIAL COUPLING

When a piece of Si with some Li in it is introduced into an electrolyte containing Li$^+$, it is expected that the following equilibrium is established at the interface between the Li atoms in silicon, Li$^+$ ions in the electrolyte, and electrons in silicon.

$$Li\,(Si) \rightleftharpoons Li^+(elyte) + e^-(Si) \qquad 2$$

where a symbol in parenthesis refers to the phase where the preceding species is found ("*elyte*" stands for electrolyte). The equilibrium condition for the above reaction can be written as

$$\sum_i v_i \tilde{\mu}_i = 0 \qquad 3$$

where $v_i$ is the stoichiometric number of each species $i$, and $\tilde{\mu}_i$ is its electrochemical potential, which is defined as

$$\tilde{\mu}_i = \mu_i + z_i F \varphi_I \qquad 4$$

where $\mu_i$ is the chemical potential of species $i$, $\varphi_I$ is the electric potential in phase $I$ in which the charged species is found, $F$ is the Faraday constant, and $z_i$ is the valence of species $i$. So, at equilibrium,

$$\tilde{\mu}_{Li}(Si) = \tilde{\mu}_{Li^+}(elyte) + \tilde{\mu}_{e^-}(Si) \qquad 5$$

Taking the Li atoms in Si to be neutral, and combining equation 5 with equation 1 gives,

$$\mu_{Li}^0(Si) + RT \log\left(\gamma_1 \frac{c}{c_{max}-c}\right) - \frac{v_{Si}\eta}{3}\sigma_{kk} - \frac{v_{Si}\beta_{ijkl}}{2}\sigma_{ij}\sigma_{kl}$$
$$= \left[\mu_{Li^+}^0(elyte) + RT \log\left(\gamma_2 \frac{m}{m_0}\right) + F\varphi_{elyte}\right] + \left[\mu_{e^-}^0(Si) + RT\log\left(\gamma_3 \frac{c_e}{c_{e_0}}\right) - F\varphi_{Si}\right] \qquad 6$$

The left hand side is the Larché-Cahn potential for Li in Li-Si solid solution. In the above equation, $m$ is the concentration of Li$^+$ in the electrolyte, and $m_o$ is its reference value; $c_e$ is the concentration of electrons in Si, and $c_{e_0}$ is its reference value; $\gamma_1$ is the activity coefficient of Li in Si, $\gamma_2$ is that of Li$^+$ in the electrolyte, $\gamma_3$ is that of electrons in Si, $\mu_{e^-}^0$ is the reference chemical potential of electrons in Si. Recognizing that the electron density in Si is effectively constant, the above equation can be rearranged to read:

$$\mu_{Li}^0(Si) - \mu_{e^-}^0(Si) + F(\varphi_{Si} - \varphi_{elyte}) + RT \log\left(\gamma_1 \frac{c}{c_{max}-c}\right) - \frac{v_{Si}\eta}{3}\sigma_{kk} - \frac{v_{Si}\beta_{ijkl}}{2}\sigma_{ij}\sigma_{kl}$$
$$= \left[\mu_{Li^+}^0(elyte) + RT \log\left(\gamma_2 \frac{m}{m_0}\right)\right] \qquad 7$$

The right hand side is the chemical potential of Li$^+$ in the electrolyte, $\mu_{Li^+}(elyte)$. Let the electric potential difference between the Si electrode and the electrolyte, $(\varphi_{Si} - \varphi_{elyte}) = E'$. Similarly, let the electric potential difference between the electrolyte and a reference Li counter electrode be $(\varphi_{elyte} - \varphi_{Li}) = E''$. Using equation 7, the equilibrium potential of the Li/Si half cell $E_o = E' + E''$ can be written as,

$$FE_0 = \mu_{Li^+}(elyte) - \mu' + FE''  \qquad 8$$

Where

$$\mu' = \mu^0_{Li}(Si) - \mu^0_{e^-}(Si) + RT\log\left(\gamma_1 \frac{c}{c_{max} - c}\right) - \frac{v_{Si}\eta}{3}\sigma_{kk} - \frac{v_{Si}\beta_{ijkl}}{2}\sigma_{ij}\sigma_{kl} \qquad 9$$

Note that $E''$ remains constant if the concentration of $Li^+$ in the electrolyte does not change. Equations 8 and 9 reveal the coupling between the state of stress $\sigma_{ij}$ in the Si electrode and the cell potential $E_o$. Traditionally, the change in the electrode potential with lithiation is thought to be due to the change in the entropic (log) term in equation 9, and possibly due to change in the binding energy, which could depend on concentration (which would be reflected in the activity coefficient). Equation 9 shows that stress can affect the potential in two ways: the last term represents the rate of change of elastic energy due to change in the elastic constants; the preceding term is analogous to the $P\Delta V$ term in the chemical potential of a gaseous mixture, where work must be done against the pressure $P$ to increase the volume by $\Delta V$ by introducing additional material.

It is instructive to examine the relative magnitudes of these two stress terms. Assuming that the elastic properties of the material remain isotropic, the compliance tensor $s_{ijkl}$, in terms of the elastic modulus $E$ and the Poisson's ratio $v$, is given by

$$s_{ijkl} = \frac{1+v}{E}\left(\delta_{il}\delta_{jk} + \delta_{ik}\delta_{jl}\right) - \frac{v}{E}\delta_{ij}\delta_{kl} \qquad 10$$

from which the last term in Equation 9 becomes

$$\frac{\beta_{ijkl}}{2}\sigma_{ij}\sigma_{kl} = \frac{\partial}{\partial c}\left(\frac{1+v}{E}\right)\sigma_{ij}\sigma_{ij} - \frac{\partial}{\partial c}\left(\frac{v}{E}\right)\sigma_{kk}^2 \qquad 11$$

In the context of the experiments on thin-film silicon electrodes presented in this paper, $\sigma_{ij}$ represents a state of equi-biaxial stress $\sigma$ in the plane of the film and zero stress in the out-of-plane direction, for which equation 11 becomes,

$$\frac{\beta_{ijkl}}{2}\sigma_{ij}\sigma_{kl} = \sigma^2 \frac{\partial}{\partial c}\left(\frac{1}{E^*}\right) \qquad 12$$

where $E^*$ is the biaxial modulus, $E^* = E/(1-v)$. The ratio ($r$) of the last term in Equation 9 to its preceding term can be expressed as

$$r = \frac{3}{2}\frac{\sigma}{\eta}\frac{\partial}{\partial c}\left(\frac{1}{E^*}\right) \qquad 13$$

In earlier experiments,[50] it was reported that the stress in silicon films during electrochemical cycling is of the order of 1 GPa. From published results on the rate of volumetric expansion of silicon during lithiation,[1,18] $\eta \sim 0.7$. In a separate experimental investigation, we measured the biaxial modulus of silicon thin films as a function of lithium concentration,[56] and found that $E^*$ varies from around 80 GPa when $c = 0.3$, to about 35 GPa when $c \sim 3$. Similar reduction in elastic modulus was predicted by the computational results of Shenoy et al.[57] From these values, $r \sim 1\%$. Hence, the contribution to the chemical potential change due to the $\beta_{ijkl}\sigma_{ij}\sigma_{kl}/2$ term is small compared to that from the $\eta\sigma_{kk}/3$ term, and will therefore be ignored in the subsequent discussion.

At fixed Li concentration in Si and Li$^+$ concentration in the electrolyte, the change in the cell potential in response to a change in stress can be written as (from equations 8 and 9),

$$\Delta E_0 = \frac{v_{Si}\eta}{3F}\Delta\sigma_{kk} \qquad 14$$

The density of amorphous silicon is reported to be 2.2 g/cm$^3$,[58] from which $v_{Si} \sim 12.7$ cm$^3$/mol and F = 96485 C/mol. From these values, the stress-potential coupling in silicon in a thin film configuration can be estimated to be

$$\frac{\Delta E_0}{\Delta\sigma} \approx 62 \; mV/GPa \qquad 15$$

In other words, a compressive stress of 1 GPa decreases the Li/Si half-cell potential by ~62 mV, and a tensile stress of 1 GPa increases this potential by the same value. In order to examine the relative importance of this number, consider the potential change during lithiation of silicon in a Si/Li half cell. For most of lithiation, the potential decreases monotonically between ~ 0.5 V and 10 mV vs. Li/Li$^+$; hence, a potential change of 60 mV is about 25% of the average lithiation potential, which implies that the stress effect can be substantial and should be taken into account in lithium-ion battery modeling and energy efficiency calculations.

In the experiments reported by Sethuraman et al.,[50] during a lithiation–delithiation cycle, the difference in compressive and tensile stresses was seen to vary around 2-3 GPa, which by the above estimate, can contribute substantially to potential offset in the hysteresis loop of the potential vs. SOC plot. The energy dissipation represented by this fraction of the hysteresis plot corresponds to the mechanical energy dissipation due to plastic flow of the film during lithiation (plastic flow in compression), and de-lithiation (plastic flow in tension). These observations motivate the present experimental investigation to measure the stress-potential coupling in silicon thin-film electrodes, which is described in the next section.

## 3. EXPERIMENTAL

*Electrochemical cell.* - Silicon wafers [single-side polished, 50.8 mm diameter, nominally 425-450 μm thick, (111) orientation, and with 200 nm thermally grown oxide on all sides] were used as substrates for electrode fabrication. The oxide layer isolates the silicon wafer from participating in the electrochemical reactions. A 500 nm copper thin film was first sputtered onto the unpolished side of the Si wafer followed by the deposition of ~250 nm silicon film (see Figure 1). Copper thin films were prepared by DC-sputtering of a copper target (2" diameter, 0.25" thick disc, 99.995%, Kurt J. Lesker Company) at 150 W and a pressure of 0.266 Pa of Argon. Silicon thin films were prepared by RF-magnetron sputtering of a silicon target (2" diameter, 0.25" thick disc, 99.995% Si) at 180 W power and at a pressure of 0.266 Pa of Argon (99.995%). Previous studies have shown that silicon sputtered under these conditions results in amorphous films.[38] Further, it was shown in literature that the Cu underlayer is critical to the cycling of Si thin films,[59,60] which serves as a current collector and aids in uniform current distribution on the Si electrode. The electrochemical-cell assembly is shown schematically in Figure 1.

The Si wafer (coated with the Cu and Si thin films) was then assembled into a home-made electrochemical cell with a glass window (Figure 1). Lithium metal acts as counter and reference electrode (diameter = 5.08 cm, thickness = 1.5 mm), which is separated from the Si electrode by a woven Celgard C480 separator (thickness = 21 μm, Celgard Inc., Charlotte, North Carolina). 1.2 M lithium hexafluoro-phosphate in 1:2 (vol. %) ethylene carbonate:diethyl carbonate with 10% fluoroethylene carbonate additive was used as the electrolyte. The FEC additive increases the cycling efficiency of silicon electrodes, probably due to the formation of a stable solid-electrolyte-interphase (SEI) layer.[61]

*Electrochemical measurements.* - Electrochemical measurements were conducted in an environmental chamber in ultra-high pure argon atmosphere at 25°C (±1°C) using a Solartron 1470E MultiStat system (Solartron Analytical, Oak Ridge, Tennessee), and data acquisition was performed using the Corrware software (Scribner Associates Inc., Southern Pines, North Carolina). The cell was cycled galvanostatically at a current of 12.5 μA/cm$^2$ (*ca.* C/8 rate) between 1.2 V and 0.01 V *vs.* Li/Li$^+$. The data acquisition rate was 1 Hz for all the electrochemical experiments. The lower limit of 0.01 V *vs.* Li/Li$^+$ was chosen to avoid lithium plating and also to avoid the formation of the crystalline Li$_{15}$Si$_4$ phase. Open-circuit relaxation experiments were conducted at the end of lithiation and delithiation steps, as well as during the stress-potential experiments. The input impedance of the instrument was 12 GΩ, and hence the current due to the open-circuit potential measurement was negligible.

*In situ stress measurements.* - Stresses in the silicon thin-film were measured by monitoring the substrate-curvature changes during electrochemical lithiation and delithiation. The relationship between the biaxial film stress, $\sigma_f$, and the substrate curvature, $\kappa$, is given by the Stoney equation,[62,63]

$$\sigma_f = \frac{E_s h_s^2 \kappa}{6 h_f (1-\upsilon_s)} \qquad 16$$

where $E_s$ and $\upsilon_s$ are the Young's modulus and the Poisson's ratio of the substrate respectively, and $h_s$ is the substrate thickness. Values for $E_s$ and $\upsilon_s$ were obtained from Brantley's work.[64] Based on *in situ* observations of height and volume changes during lithiation/delithiation in thin-film silicon electrodes[65,66] reported in the literature, it is reasonable to assume that the film thickness increases linearly with the state of charge (SOC), i.e., $h_f = h_f^0 (1+2.7\,z)$, where $h_f^0$ is the initial film thickness, and $z$ is the state of charge, which can vary between 0 and 1. Here, $z = 1$ corresponds to a charge capacity of 3579 mAh/g, which corresponds to a volume expansion of 370%.

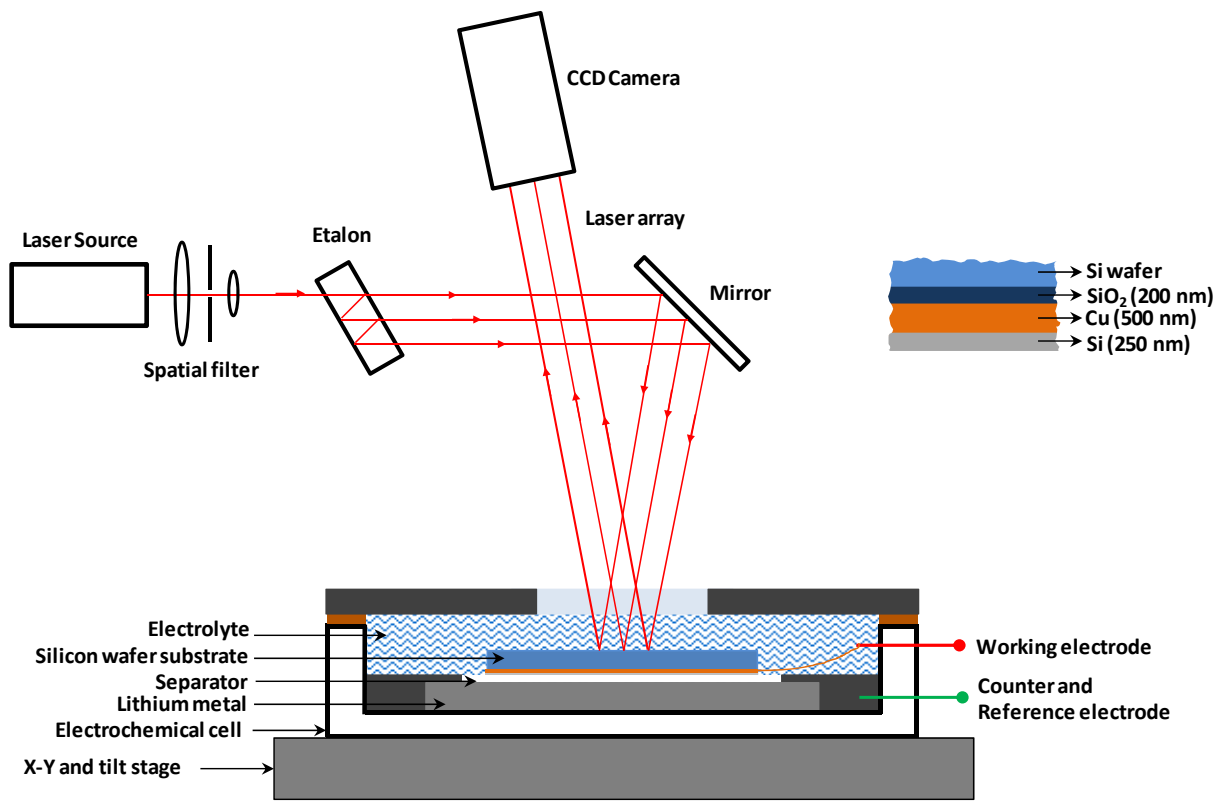

*Figure 1: Schematic illustration of the home-made electrochemical cell is shown along with the multi-beam-optical sensor setup to measure substrate curvature. Note that the schematic is not drawn to scale. In this two-electrode configuration, Si thin film is the working electrode, and Li metal is the counter and reference electrode. Above right: the layered configuration of the working electrode on Si-wafer substrate is shown.*

Substrate curvature was monitored with a multi-beam optical sensor (MOS) wafer curvature system (kSA-MOS, K-Space Associates, Inc., Dexter, Michigan), which is illustrated schematically with the electrochemical cell in Figure 1. The MOS system uses a parallel array of laser beams that get reflected off the sample surface and captured on a CCD camera. The relative change in the spot spacing is related to the wafer curvature through

$$\kappa = \frac{(d - d^0)}{d^0} \frac{1}{A_m} \qquad 17$$

where $d$ is the distance between two adjacent laser spots on the CCD camera (see figure 1(b) in reference 50), $d_0$ is the initial distance and $A_m$ is the mirror constant, given by $2L/\cos(\theta)$; $L$ is the optical path length of the laser beam between the sample and the CCD array and $\theta$ is the incident angle of the laser beam on the sample. The mirror constant $A_m$ is measured by placing a reference mirror of known curvature in the sample plane and measuring the relative change in the spot spacing. Spot spacing data was collected at a rate of 1 Hz in all experiments. Throughout this study, the sign convention for compressive stress is negative while that for tensile stress is positive.

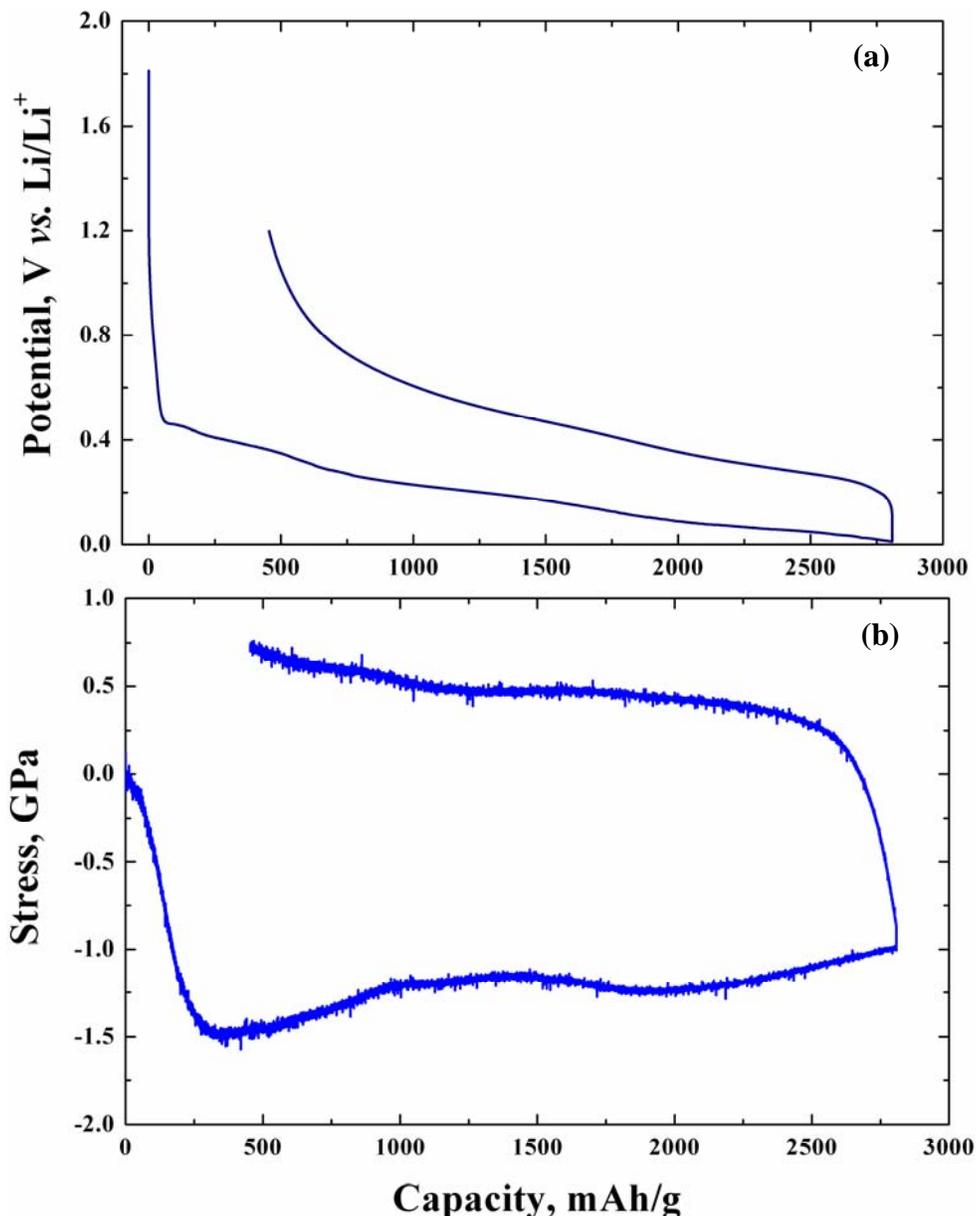

*Figure 2: (a) Transient cell potential corresponding to the initial lithiation and delithiation of magnetron-sputtered amorphous silicon thin-film electrode cycled at C/8 rate between 1.2 and 0.01 V vs. Li/Li$^+$, and (b) the corresponding stress transience calculated from the substrate-curvature data using the Stoney equation.*

Figure 2 shows variation of potential and stress during a lithiation-delithiation cycle. Upon lithiation, the substrate constraint results in compressive stress in the film, the magnitude of which increases linearly with capacity. The rapid, linear increase in the compressive stress indicates initial elastic response. At a compressive stress of about 1.5 GPa, the film appears to reach the elastic limit and the stress reaches a plateau, which indicates that further volume expansion is accommodated by plastic deformation in the film. The flow stress is seen to generally decrease with lithiation. Note that the actual stress magnitude could be somewhat different from the values mentioned above, offset by the initial residual stress induced during the sputtering process (which was not measured in this investigation). However, for the purposes of this study, as seen in equation 14, the relevant quantity is the stress change, not the absolute value of the stress.

Upon delithiation, the unloading (decrease in stress magnitude) is initially elastic; the stress reverses elastically until it reaches a tensile plateau, where the film begins to plastically deform in tension in order to accommodate further reduction in volume. Thus, the film undergoes repeated compressive and tensile plastic flow during successive lithiation-delithiation processes, respectively. Note that, when delithiation starts, since the unloading is elastic, it takes a very small change in SOC for the stress to change from the compressive flow value to the tensile flow value, *i.e.*, a stress change of more than 1 GPa occurs for a mere ~ 2% change in SOC. In other words, by proceeding to delithiate in small increments, the stress can be changed by more than 1 GPa, while keeping the SOC practically constant, which allows measurement of change in cell potential ($\Delta E_o$) as a function of the change in stress ($\Delta \sigma$) at an almost constant SOC. This is the basis for the measurements reported in this paper. Moreover, from the slope of the potential *vs*. SOC plot just preceding delithiation (where the compressive stress is approximately constant), it is possible to correct the $\Delta E_o$ *vs*. $\Delta \sigma$ measurement for the small change in SOC that occurs during the measurement.

*Coupled stress-potential ($\sigma$-$E_o$) measurements.* - At a given SOC, the stress-potential dependence in a lithiated-silicon electrode was measured as follows: the electrode was first lithiated to a desired SOC or potential, $P$ ($1.2 \leq P \leq 0.10$ V *vs*. Li/Li$^+$), then delithiated for a small period of time, $t_d$, followed by an open-circuit-potential relaxation for one hour. From mechanics standpoint, as discussed above, the delithiation step is equivalent to elastic unloading. During the hour-long open circuit, the potential relaxes, presumably towards its equilibrium value; similarly, the stress also relaxes towards a steady-state value. As can be seen from Figure 3, although the rates of relaxation of the potential and the stress become very small at the end of one hour, true equilibrium may not be reached even after a long time because of the side reactions that continue even in the absence of external current. Hence, the steady state value of the potential at the end of the hour-long open-circuit step are taken to be representative of the values predicted by Eq. 8 for the system at that SOC. Note that the potential measured is a mixed potential due to the presence of the SEI side reaction that occurs simultaneously with the lithiation/delithiation reaction.[44] While this can be accounted for by correcting for the side reaction,[39,44] in this paper our analysis deals with changes in the potential as opposed to the absolute value, as described below, and hence this correction is not necessary.

After the open circuit relaxation step, the electrode was re-lithiated to the same initial potential *P*, and delithiated for an incrementally larger duration than before, *i.e.*, for a period ($t_d + \delta_t$), followed by an hour-long open-circuit relaxation. The increase in delithiation time allows for incrementally larger elastic unloading thereby taking the electrode stress to a new value. This sequence of steps was repeated with progressively longer delithiation time, allowing the sample to relax towards a steady state potential at several values of stress between the compressive and tensile flow stress, resulting in a total stress change ($\Delta\sigma$) of over 1 GPa.

As noted above, progressive delithiation changes the film stress from compression to tension; the latter can lead to film cracking when the stress magnitude exceeds a critical value (the value of which is determined by the fracture toughness of the film, its thickness and initial defect distribution). When the crack density is sufficiently high, the relation between the film stress and the substrate curvature should be corrected for the crack density. Hence, the coupling between stress and potential is best measured during first lithiation, when the stress is known to be compressive and cracks do not form. During the later stage of delithiation steps, when the stress is tensile and sufficiently high, cracks are likely to form and the procedure employed in this paper (which does not account for crack density) would not be exact in that regime. Hence, attention should be focused primarily on data corresponding to states of compressive stress.

## 4. RESULTS AND DISCUSSION

Figure 3a-c shows the history of current, potential and stress during several delithiation-relithiation steps in a stress-potential experiment. At the end of the first few open-circuit steps, the potential and stress appear to be still evolving;[38] however, in the later steps, the rate of change at the end of open circuit relaxation was negligible (Figure 3b). Note that, for the sake of clarity, Figure 3 shows only the first few delithiation-relithiation steps in a long sequence that spans a stress change of over 1 GPa. The measured potential $E_o$ at the end of the one-hour relaxation period is plotted against the corresponding silicon electrode stress in Figure 4a, for three different experiments in which the initial capacity values are 463, 2393 and 3441 mAh/g. Also shown on the same plot are dotted lines with a slope of 62 mV/GPa, which is the value predicted by the analysis in section 2.

For a more direct comparison with equation 14, the potential change ($\Delta E_o$) is plotted as a function of the change in stress ($\Delta\sigma$) in Figure 4b. The stress and potential changes are measured with respect to the initial values at which progressive delithiation begins (the figure also shows a dotted line also with a slope of 62 mV/GPa). The experimentally measured slopes corresponding to the three initial states of charge of 0.9, 0.67 and 0.13 are 110 mV/GPa, 104 mV/GPa and 125 mV/GPa, respectively. These numbers indicate a significant coupling between the electrode stress and the electrochemical performance of silicon electrodes, which needs to be taken into account in designing silicon based electrodes and in evaluating their performance. It is encouraging that the predicted and measured values are of the same order of magnitude, although the former is 40 – 50% smaller than the latter. The agreement can be considered to be reasonable, given the limitations of the Larché and Cahn potential discussed in section 2. The difference could also imply that the physics of the problem is not being captured fully by the Larché-Cahn potential.

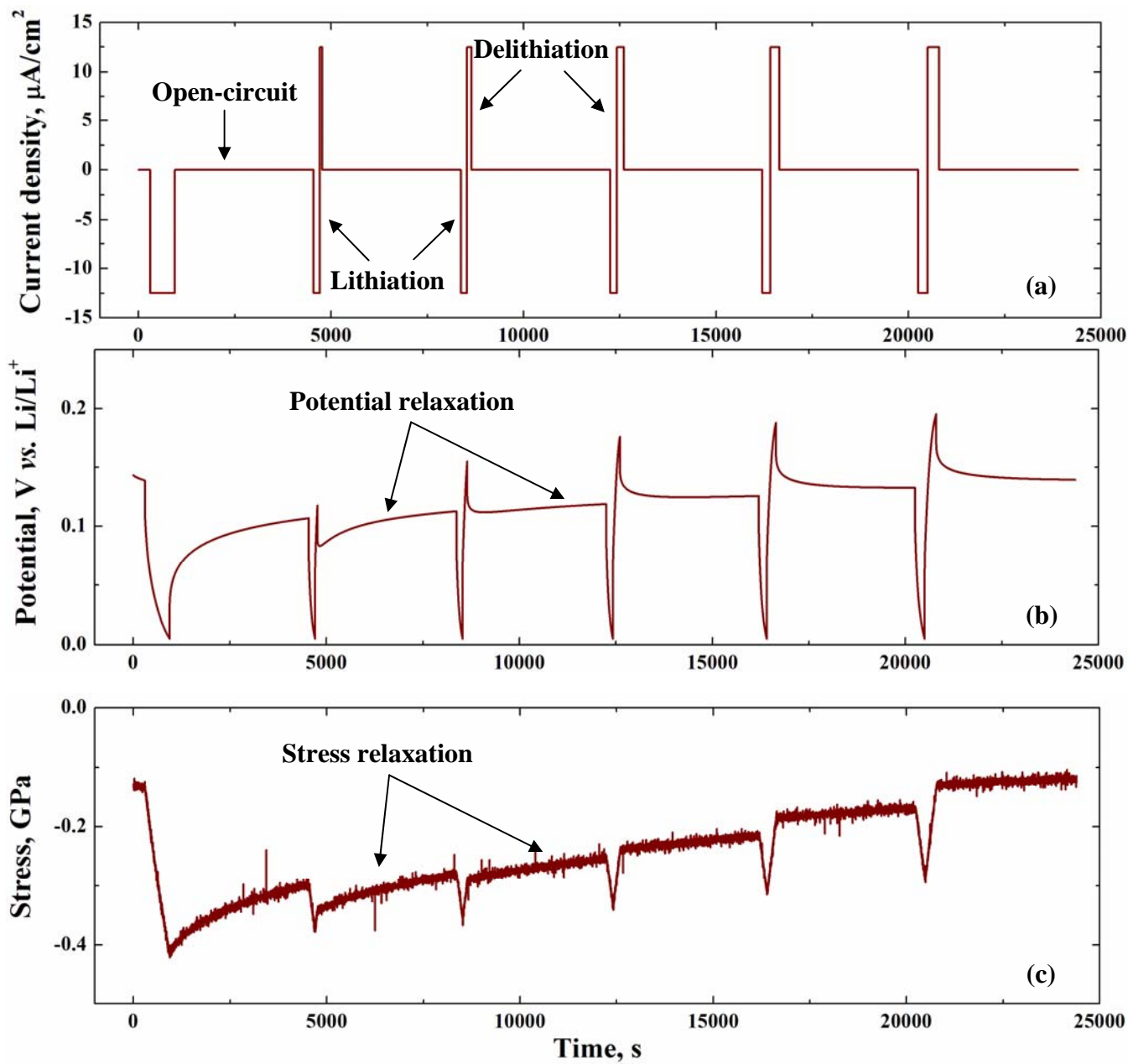

*Figure 3: Representative current density, potential and stress transience obtained from the stress-potential experiment in a lithiated-silicon electrode. These experiments were done on a silicon electrode lithiated to a capacity of 3441 mAh/g (SOC ≈ 0.9). The cutoff potential, P, for this experiment was 0.01 V vs. Li/Li$^+$, initial delithiation time, $t_d$, was 60 seconds, and the incremental increase in delithiation time, $\delta_t$, thereafter was 60 seconds for each delithiation experiment.*

As discussed in the previous section, the SOC changes by about 1.5% between the beginning and the end of the incremental delithiation, while the stress varies by more than 1 GPa (ideally, the stress should be changed without changing the SOC). The measured potential change can be corrected for this small change in the SOC by examining the variation of the potential and the stress as a function of SOC just prior to the delithiation steps, as shown in Figure 5. According to the analysis presented in section 2, equations 8 and 9 represent the variation of the potential due to change in Li concentration and stress in the silicon film. However, since the rate of change in stress at point C (where delithiation starts) is small (it can be seen from Figure 2b that the stress changes are rather benign after the initial steep increase) the change in potential is essentially due to the change in the concentration term in equation 9. Just prior to point C, the change in potential with Li concentration can be approximated as being linear, with a slope of -0.121 mV/mAhg$^{-1}$. For the case shown in Figure 4, the cumulative change in SOC during all delithiation steps is ~50 mAh/g, which translates a potential change of ~6 mV due to the change in SOC. The data shown in Figure 4 incorporates this correction. Note that the correction is only about 5% of the measured change in potential.

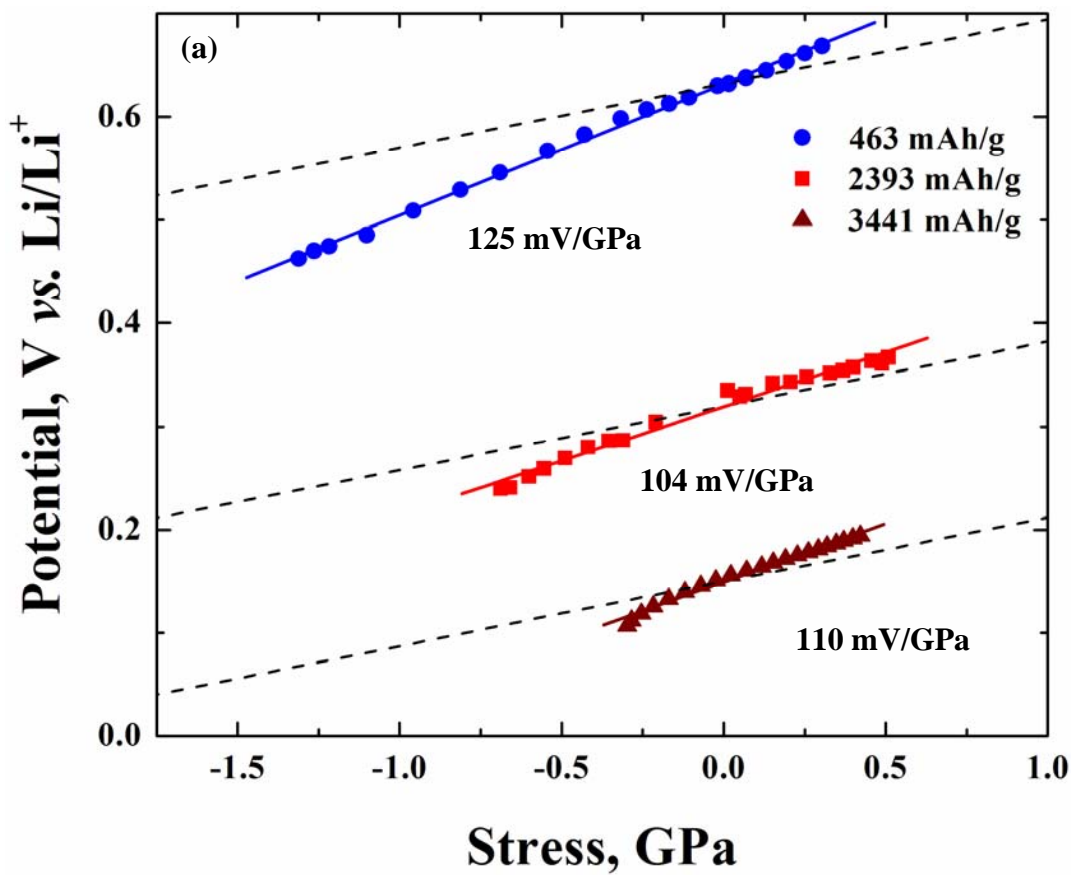

*Figure 4a: Potential, and stress values at the end of the open-circuit-potential relaxation period in the stress-potential experiments conducted at three different SOCs are shown along with linear fits (solid lines). Potential is corrected for the change in Li concentration due to the incremental delithiation steps in the stress-potential experiments. The dashed line for each SOC represents the Larché-Cahn potential centered around the Y-intercept of the linear fit for that SOC.*

Note that the *y*-intercept in Figure 4a can be taken as a measure of equilibrium potential of lithiated silicon at a state of zero stress; for the SOCs of 0.13, 0.67 and 0.9, the respective *y*- intercepts are 0.632 V, 0.32 V and 0.15 V, all *vs.* Li/Li$^+$. However, these values should be corrected for the initial residual stresses to arrive at true equilibrium potential that excludes stress effects, which will be pursued in future work.

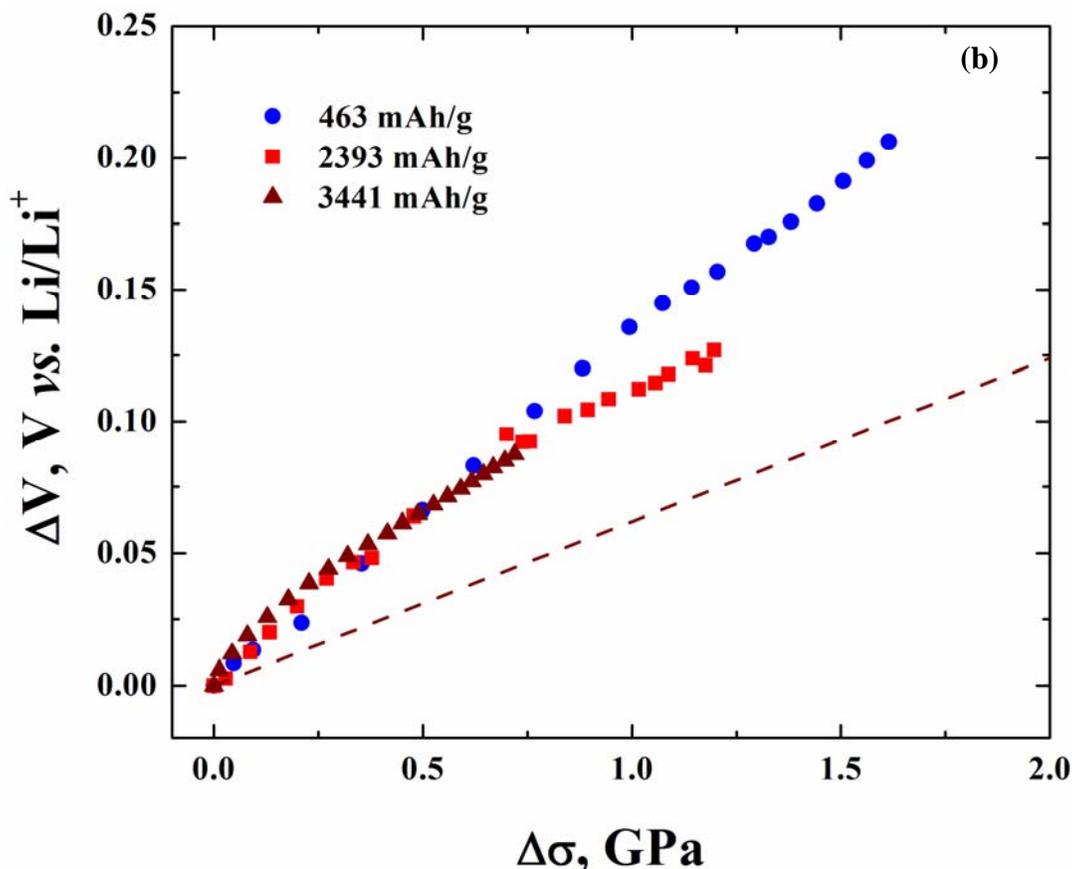

*Figure 4b: Normalized potential, and stress values at the end of the open-circuit-potential relaxation period in the stress-potential experiments is shown with the Larché-Cahn potential (dashed line).*

Note that the evolution of the potential reported in Figure 3b can also be a consequence of a sluggish reaction[38,39] which results in a large time constant to reach equilibrium combined with a side reaction that self-discharges the electrode and results in a mixed potential. Accordingly, the potential reported in Figures 4a and b would need to be corrected for this effect. However, the purpose of this paper is to illustrate the coupling between potential and stress, as we have demonstrated using theory and the experimental evidence. The coupling of these various effects is outside the scope of this paper, and will be explored in the future. While the existence of stress-potential coupling has been demonstrated in thin-film geometry, such measurements have to be made on composite electrodes that are of relevance to battery industry. *In situ* stress measurements on commercial battery electrodes are currently ongoing in our laboratory and results from these studies will be reported in the future.

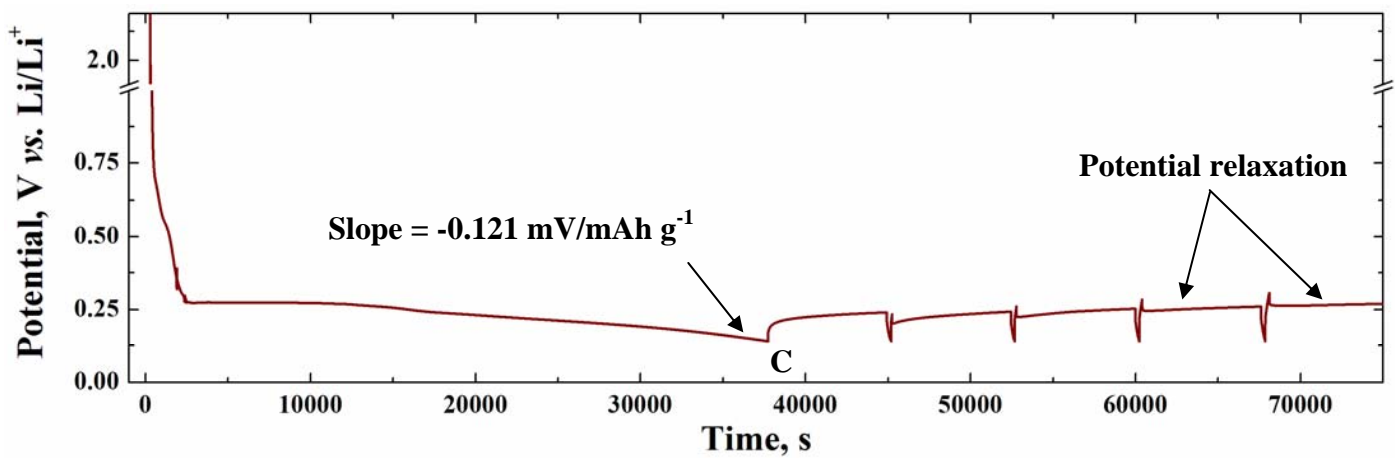

*Figure 5: Potential vs. time plot during lithiation – incremental delithiation of silicon electrode. The electrode is lithiated continuously until point C, where the delithiation steps begin, as described in the text. The slope of the potential – SOC relation just prior to point C is 0.121 mV/mAhg$^{-1}$, which is used to correct for the change in SOC during the delithiation steps.*

The magnitude of the stress-potential coupling reported here has several implications. For example, using Galvanostatic Intermittent Titration Technique (GITT), Ryu et al.[67] and Baggetto et al.[46] have reported the equilibrium potential as a function of SOC for the lithiated-silicon system. Similar data have been reported by Ding et al. recently.[68] For every SOC value, they report two equilibrium potentials, one corresponding to lithiation direction and one corresponding to the delithiation direction. However, interpretation of such data may have to consider the stress effect as well because, according to equations 8 and 9, compressive stress and tensile stress respectively lowers and raises the equilibrium potential *vs.* SOC plot. Hence, the effect of stress may be important in interpreting GITT measurement of equilibrium potential in lithiated-silicon electrodes. At a given SOC, difference in the equilibrium potential values (corresponding to lithiation and delithiation) reported by Baggetto et al. varies from as 280 mV at $Li_{1.25}Si$ to 150 mV at $Li_{3.5}Si$. These values correspond to approximately 50-60% of the total potential offset between lithiation and delithiation at a C/10 rate reported in their study. They hypothesize that the origin of the observed potential offset between lithiation and delithiation at a given SOC could be thermodynamic in nature. In addition to this hypothesis proposed by Baggetto et al. to explain different values of the equilibrium potential during lithiation and delithiation, it is possible that the difference in stress between the two processes (which can be as high as 2 GPa as reported here and elsewhere)[50] also contributes to the difference.

The stress effect on the potential has implications in interpreting the potential hysteresis loops observed during delithiation-lithiation cycles; the area of the loop represents the energy dissipation in each cycle. From the foregoing discussion, it is clear that the potential offset exhibited by lithiated-silicon has two components: the overpotential required to drive a finite current and the stress effect. In other words, consider the equilibrium potential *vs.* SOC relation for lithiated silicon in the absence of any stress, which also represents the result of a hypothetical measurement at an infinitesimally small current in an unconstrained Si film. However, during lithiation at a finite current (still stress free), the measured potential *vs.* SOC relation would be lower than the equilibrium relation by the overpotential required to drive the current. Similarly, during delithiation, the measured potential would be higher than the equilibrium value by the same overpotential. The area of the resulting hysteresis loop represents the energy dissipation due to polarization losses (*i.e.,* the sum of kinetic, ohmic and concentration overpotentials). Now, consider the lithiation of a thin film on a substrate in our experiment, which develops compressive stress. From equation 9, the compressive stress lowers the potential (equation 14). During delithiation, since the stress is tensile, the potential is raised above the equilibrium value, by an amount determined by the magnitude of the tensile stress. Thus, the combination of compressive and tensile stress during lithiation and delithiation respectively makes an additional contribution to the area of the hysteresis loop, which represents the mechanical energy dissipation due to plastic deformation of lithiated silicon in both halves of the cycle. If the electrode deformation were purely elastic, the measured potential would still deviate from the equilibrium value, but there would be no contribution to the hysteresis loop from the stress effect; lithiation and delithiation in that case would represent loading and unloading along the same stress path. In experiments reported earlier,[50] it was shown that the energy dissipation due to plastic deformation of lithiated silicon can be comparable to the polarization losses. Eliminating the mechanical dissipation would result in an equivalent improvement of energy efficiency of the system. Hence, stress considerations suggest that silicon electrodes should be designed to minimize or eliminate plastic deformation.

Electrode stress has implications for the maximum realizable charge/discharge capacity as well, very similar to the way in which the kinetic overpotential and ohmic drop affect these capacities. As discussed above, a compressive stress of ~ 1 GPa depresses the potential of the silicon electrode by more than 60 mV, and a tensile stress of ~ 1 GPa elevates it by a similar value. As a result, during lithiation, the lower cut-off potential (typically set at 10 mV *vs.* Li/Li$^+$) is reached at a lower value of SOC. Note that most of the lithiation takes place below a potential of 500 mV; the displacement of the potential curve due to stress is a significant fraction of 500 mV, hence the reduction in the charge capacity at the lower cut-off potential can be substantial. Similarly, during delithiation, the upper cut-off potential is usually set at 1.2 V *vs.* Li/Li$^+$; because of the tensile stress during delithiation, which elevates the potential, the upper cut-off potential is reached at a higher SOC, resulting in a reduction in the maximum realizable discharge capacity. Thus, silicon electrode designs that minimize or eliminate stresses have the additional advantage that their charge/discharge capacities will be higher (*i.e.,* increased electrode utilization) at higher rates.

## 5. CONCLUSIONS

A thermodynamic argument based on the Larché-Cahn chemical potential is made to demonstrate the existence of stress-potential dependence in a lithiated-silicon system. From the known properties of silicon, it is estimated that the ratio of potential change to stress change would be about 62 mV/GPa. An experimental investigation has been carried out, in which the stress dependence of the potential was measured at three different states of charge, and the ratio of potential change to stress change was measured to be in the range of 100 – 125 mV/GPa, which can be considered to be a reasonable agreement, considering the limitations of the Larché-Cahn potential. It is also likely that the difference between the experimental measurements and the thermodynamic prediction is a consequence of treating lithiated silicon as a simple amorphous solid solution during lithiation and delithiation, which may not be a true representation of the actual changes in the material and its structure; these issues requires further study. The measured potential shift due to stress is a significant fraction of the average lithiation potential; hence, stress in silicon electrodes can influence their charge/discharge capacity, energy dissipation due to plastic deformation and interpretation of GITT measurements of equilibrium potential.

**Table 1: Parameters used for the stress-potential calculations presented in this study.**

| Parameter | Definition | Value | Comments |
|---|---|---|---|
| $d_f$ | Film diameter (also Si wafer diameter) | 5.08 cm | Measured |
| $E_s$ | Young's modulus of Si (111) wafer | 169 GPa | Ref. 64 |
| $c_{max}$ | Maximum stoichiometric ratio in lithiated silicon system | 3.75 | Ref. 23 |
| $(d\varepsilon/dc)$ | Linear strain of lithiated silicon | 0.146 | Ref. 23 |
| $h_f^0$ | Initial film thickness | 250 nm | Measured |
| $h_s$ | Wafer thickness | 450 μm | Measured |
| $2L/\cos(\theta)$ | Mirror constant | 2.96 m | Measured |
| $\eta$ | Rate of volume expansion | 0.7 | Ref. 1,18 |
| $v_s$ | Poisson's ratio of Si (111) | 0.26 | Ref. 64 |
| $v_{Si}$ | Molar volume of amorphous Si | 12.7 cm$^3$/mol | Ref. 58 |
| $\rho_f$ | Silicon film density | 2.2 g/cm$^3$ | Ref. 58 |

## 6. ACKNOWLEDGEMENTS

The authors gratefully acknowledge support of this work by the Materials Research, Science and Engineering Center (MRSEC) sponsored by the United States National Science Foundation under contract no. DMR0520651; and by the Rhode Island Science and Technology Advisory Council under grant no. RIRA 2010-26. V.S. gratefully acknowledges support from the Assistant Secretary for Energy Efficiency, United States Department of Energy, under contract no. DE-AC02-05CH11231. The authors thank Celgard Inc. for providing high-rate separator samples.

## 7. LIST OF SYMBOLS

$A_m$     Mirror constant, m
$c$     Concentration of Li; ratio of the number of Li atoms to that of Si in a unit volume
$c_e$     Concentration of electrons in Si
$c_{e_0}$     Reference concentration of electrons in Si
$c_0$     Reference concentration of Li
$d$     Distance between adjacent laser spots on the CCD camera
$d_f$     Diameter of the silicon film, m

| | |
|---|---|
| $d^0$ | Initial distance between adjacent laser spots on the CCD camera |
| $e^-$ | Electron |
| $E$ | Elastic modulus, GPa |
| $E_0$ | Equilibrium potential of Li/Si half cell, V |
| $E'$ | Potential difference between Si and electrolyte, $(\varphi_{Si} - \varphi_{elyte})$, V |
| $E''$ | Potential difference between electrolyte and Li, $(\varphi_{elyte} - \varphi_{Li})$, V |
| $E^*$ | Biaxial modulus, $E/(1-v)$, GPa |
| $E_s$ | Young's modulus of the silicon-wafer substrate, GPa |
| $F$ | Faraday constant, 96485 C mol$^{-1}$ |
| $h_f$ | Film thickness, m |
| $h_f^0$ | Initial film thickness, m |
| $h_s$ | Substrate thickness, m |
| $L$ | Optical path length of the laser beam |
| $m$ | Concentration of Li$^+$ in electrolyte |
| $m_0$ | Reference concentration of Li$^+$ in electrolyte |
| $P$ | Cut-off potential *vs.* Li/Li$^+$ in the stress-potential experiment, V |
| $R$ | Universal gas constant, 8.314 J mol$^{-1}$ K$^{-1}$ |
| $T$ | Absolute temperature, K |
| $t_d$ | Delithiation time in the stress-potential experiment, s |
| $s_{ijkl}$ | Compliance tensor |
| $v_{Si}$ | Molar volume of Si, m$^3$ mol$^{-1}$ |
| $z$ | State of charge, SOC |
| $z_e$ | State of charge at the end of elastic range |
| $z_i$ | Valence of species $i$ |

*Greek symbols*

| | |
|---|---|
| $\beta_{ijkl}$ | Rate of change of compliance tensor with Li concentration, $ds_{ijkl}/dc$ |
| $\gamma_1$ | Activity coefficient of Li in Si |
| $\gamma_2$ | Activity coefficient of Li$^+$ in the electrolyte |
| $\gamma_3$ | Activity coefficient of electrons in Si |
| $\delta_{ij}$ | Kronecker delta |
| $\delta_t$ | Incremental delithiation time in the stress-potential experiments, s |
| $\varepsilon_{ij}$ | Strain tensor |
| $\eta$ | Rate of change of volumetric strain of Li$_x$Si, $d\varepsilon_v^c/dc$ |
| $\theta$ | Incident angle of the laser beam on the substrate |
| $\kappa$ | Substrate curvature, m$^{-1}$ |
| $\mu_{Li}$ | Chemical potential of Li |
| $\tilde{\mu}_i$ | Electrochemical potential of species $i$ |
| $\mu_{Li}^0$ | Reference chemical potential of Li |
| $\mu_{e^-}^0$ | Reference chemical potential of electrons in Si |
| $\sigma$ | Equi-biaxial stress, GPa |
| $\sigma_f$ | Equi-biaxial film stress, GPa |
| $\sigma_{ij}$ | Stress tensor |
| $v$ | Poisson's ratio |
| $\varphi_I$ | Electric potential of phase $I$ |

*Subscripts, superscripts*

| | |
|---|---|
| *0* | Reference |
| *c* | Concentration, compositional |
| *e* | Elastic |
| *elyte* | Electrolyte |
| *f* | Film |
| *Li* | Lithium |
| *p* | Plastic |
| *s* | Substrate |

## 8. REFERENCES

1. M.N. Obrovac and L. Christensen, *Electrochem. Solid State Letters,* **7**, A93 (2004).
2. J. Li and J.R. Dahn, *J. Electrochem. Soc.,* **154**, A156 (2007).
[3]. J.-M. Tarascon and M. Armand, *Nature*, **414**, 359 (2001).
4. V.A. Sethuraman, L.J. Hardwick, V. Srinivasan, and R. Kostecki, *J. Power Sources*, **195**, 3655 (2010).
5. Specific energy on a volumetric basis, after accounting for volume expansion.
6. U. Kasavajjula, C. Wang, and A.J. Appleby, *J. Power Sources*, **163**, 1003 (2007), and references therein.
7. C.K. Chan, H. Peng, G. Liu, K. McIlwrath, X.F. Zhang, R.A. Huggins, and Y. Cui, *Nature Nanotech.,* **3,** 31 (2008).
8. B. Gao, S. Sinha, L. Fleming, and O. Zhou, *Adv. Mater.,* **13**, 816 (2001).
9. K. Peng, J. Jie, W. Zhang, and S.-T. Lee, *Appl. Phys. Lett.,* **93,** 033105 (2008).
10. M. Green, E. Fielder, B. Scrosati, M. Wachtler, and J.S. Moreno, *Electrochem. Solid State Lett.*, **6**, A75 (2003).
11. S. Bourderau, T. Brousse, and D.M. Schleich, *J. Power Sources*, **81-82**, 233 (1999).
12. A. Netz, R.A. Huggins, and W. Weppner, *J. Power Sources*, **119-121**, 95 (2003).
13. J.P. Maranchi, A.F. Hepp, and P.N. Kumta, *Electrochem. Solid State Letters,* **6**, A198 (2003).
14. V. Baranchugov, E. Markevich, E. Pollak, G. Salitra, and D. Aurbach, *Electrochem. Comm.,* **9**, 796 (2007).
15. P. Limthongkul, Y.-I. Jang, N.J. Dudney, and Y.-M. Chiang, *Acta Materialia*, **51**, 1103 (2003).
16. B. Bang, M.-H. Kim, H.-S. Moon, Y.-K. Lee, J.-W. Park, *J. Power Sources*, **156**, 604 (2004).
17. J. Graetz, C. C. Ahn, R. Yazami, and B. Fultz, *Electrochem. Solid State Letters*, **6**, A194 (2003).
18. M.N. Obrovac and L.J. Krause, *J. Electrochem. Soc.,* **154**, A103 (2007).
19. J.H. Ryu, J.W. Kim, Y.-E. Sung, and S.M. Oh, *Electrochem. Solid State Letters,* **7**, A306 (2004).
20. B.A. Boukamp, G.C. Lesh, and R.A. Huggins, *J. Electrochem. Soc.*, **128**, 725 (1981).
21. W.-R. Liu, M.-H. Yang, H.-C. Wu, S.M. Chiao, and N.-L. Wu, *Electrochem. Solid State Letters*, **8**, A100 (2005).
22. N. Dimov and M. Yoshio, *J. Power Sources*, **174**, 607 (2007).

23. M.N. Obrovac, L. Christensen, D.B. Le, and J.R. Dahn, *J. Electrochem. Soc.*, 154, A849 (2007).
24. L.Y. Beaulieu, K.C. Hewitt, R.L. Turner, A. Bonakdarpour, A.A. Abdo, L. Christensen, K.W. Eberman, L.J. Krause, and J.R. Dahn, *J. Electrochem. Soc.*, **150**, A149 (2003).
25. C.-H. Doh, N. Kalaiselvi, C.-W. Park, B.S. Jin, S.-I. Moon, and M.-S. Yun, *Electrochem. Comm.*, **6**, 965 (2004).
26. H. Dong, R.X. Feng, X.P. Ai, Y.L. Cao, and H.X. Yang, *Electrochim. Acta*, **49**, 5217 (2004).
27. H. Dong, X.P. Ai, and H.X. Yang, *Electrochem. Comm.*, **5**, 952 (2003).
28. S.-J. Lee, H.-Y. Lee, H.-K. Baik, S.-M. Lee, *J. Power Sources*, **119-121**, 113 (2003).
29. M.K. Datta and P.N. Kumta, *J. Power Sources*, **158**, 557 (2006).
30. M.K. Datta and P.N. Kumta, *J. Power Sources*, **165**, 368 (2007).
31. N. Dimov, S. Kugino, and M. Yoshio, *Electrochim. Acta*, **48**, 1579 (2003).
32. N. Dimov, K. Fukuda, T. Umeno, S. Kugino, and M. Yoshio, *J. Power Sources*, **114**, 88 (2003).
33. N. Dimov, S. Kugino, and M. Yoshio, *J. Power Sources*, **136**, 108 (2004).
34. T. Hasegawa, S.R. Mukai, Y. Shirato, and H. Tamon, *Carbon*, **42**, 2573 (2004).
35. M. Holzapfel, H. Buqa, W. Scheifele, P. Novák, and F.-M. Petrat, *Chem. Comm.*, 1566 (2005).
36. Panasonic's New Li-ion Batteries Use Si Anode for 30% Higher Capacity, Nikkei Electronics Asia, March 2010.
37. K. Yasuda, "Advanced Silicon Anode Technology for High Performance Li-ion Batteries", 5[th] International Symposium on Large Lithium Ion Battery Technology and Applications, June 9, 2009, Long Beach, California.
38. V. Srinivasan, V.A. Sethuraman, and J. Newman, *Meet. Abstr. – Electrochem. Soc.*, **802**, 607 (2008)].
39. V. Srinivasan, V. Sethuraman, J. Newman, United States Department of Energy - Office of Vehicle Technologies Annual Merit Review, Bethesda, Maryland, February 28, 2008.
40. V.A. Sethuraman, V. Srinivasan, and J. Newman, *J. Electrochem. Soc.,* to be submitted (2010).
41. S. Ohara, J. Suzuki, K. Sekine, and T. Takamura, *Electrochemistry* , **71**, 1126 (2003).
42. W.-R. Liu, Z.-Z. Guo, W.-S. Young, D.-T. Shieh, H.-C. Wu, M.-H. Yang, and N.-L. Wu, *J. Power Sources*, **140**, 139 (2005).
43. T.D. Hatchard and J.R. Dahn, *J. Electrochem. Soc.,* **151**, A838 (2004).
44. V. Srinivasan, J.W. Weidner, and J. Newman, *J. Electrochem. Soc.,* **148**, A969 (2001).
45. V.A. Sethuraman, V. Srinivasan, and J. Newman, *Meet. Abstr. – Electrochem. Soc.,* **802**, 1189 (2008)].
46. L. Baggetto, R.A.H. Niessen, F. Roozeboom, and P.H.L. Notten, *Adv. Func. Mater.*, **18**, 1057 (2008).
47. V.L. Chevrier, and J.R. Dahn, *J. Electrochem. Soc.*, **156**, A454 (2009).
48. J. Christensen and J. Newman, *J. Solid State Electrochem.*, **10**, 293 (2006).
49. J. Christensen and J. Newman, *J. Electrochem. Soc.*, **153**, A1019 (2006).
50. V.A. Sethuraman, M.J. Chon, M. Shimshak, V. Srinivasan, and P.R. Guduru, *J. Power Sources*, **195**, 5062 (2010).
51. F. Larché and J.W. Cahn, *Acta Metallurgica*, **21**, 1051 (1973).
52. F. Larché and J.W. Cahn, *Acta Metallurgica*, **26**, 53 (1978).

53. F.C. Larché and J.W. Cahn, *Acta Metallurgica*, **30**, 1835 (1982).
54. F.C. Larché and J.W. Cahn, *Acta Metallurgica*, **33**, 331 (1985).
55. S.D. Beattie, D. Larcher, M. Morcrette, B. Simon, and J.-M. Tarascon, *J. Electrochem. Soc.*, **155**, A158 (2008).
56. V.A. Sethuraman, M. Shimshak, M.J. Chon, N. Van Winkle, P.R. Guduru, Manuscript submitted to *Electrochemistry Communication*, July 30, 2010.
57. V.B. Shenoy, P. Johari, Y. Qi, *J. Power Sources*, **195**, 6825 (2010).
58. O. Renner and J. Zemek, *Czech J. Phys.*, **23**, 1273 (1973).
59. J.P. Maranchi, A.F. Hepp, A.G. Evans, N.T. Nuhfer, P.N. Kumta, *J. Electrochem. Soc.*, **153**, A1246 (2006).
60. V.A. Sethuraman, K. Kowolik, and V. Srinivasan, In Press, *J. Power Sources*, (2010).
61. N.-S. Choi, K.H. Yew, K.Y. Lee, M. Sung, H. Kim, and S.-S. Kim, *J. Power Sources*, **161**, 1254 (2006).
62. G.G. Stoney, *Proc. R. Soc. (London) A*, **82**, 172 (1909).
63. L.B. Freund and S. Suresh, Thin Film Materials: Stress, Defect Formation and Surface Evolution. Cambridge University Press, Cambridge, UK (2004).
64. W.A. Brantley, *J. Appl. Phys.*, **44**, 534 (1977).
65. L.Y. Beaulieu, T.D. Hatchard, A. Bonakdarpour, M.D. Fleischauer, and J.R. Dahn, *J. Electrochem. Soc.*, **150**, A1457 (2003).
66. L.Y. Beaulieu, V.K. Cumyn, K.W. Eberman, L.J. Krause, and J.R. Dahn, *Rev. Sci. Instr.*, **72**, 3313 (2001).
67. J.H. Ryu, J.W. Kim, Y.-E. Sung, and S.M. Oh, *Electrochem. Solid State Lett.*, **7,** A306 (2004).
68. N. Ding, J. Xu, Y.X. Yao, G. Wegner, X. Fang, C.H. Chen, and I. Lieberwirth, *Solid State Ionics*, **180** 222 (2008).